\documentclass[aps,prl,preprint,showpacs,byrevtex,amsmath,amssymb,superscriptaddress]{revtex4}

\usepackage{graphicx}
\usepackage{dcolumn}
\usepackage{bm}

\begin{document}

\title{Giant-diamagnetic and magnetization-step effects in HgMnTe monocrystal}

\author{Liangqing Zhu}
\email[Electronic mail: ]{lqzhu@yahoo.cn}
\affiliation{National Laboratory for Infrared Physics, Shanghai
    Institute of Technical Physics, Chinese Academy of Sciences, 200083
    Shanghai, China}

\author{Tie Lin}
\affiliation{National Laboratory for Infrared Physics, Shanghai
    Institute of Technical Physics, Chinese Academy of Sciences, 200083
    Shanghai, China}

\author{Jun Shao}
\affiliation{National Laboratory for Infrared Physics, Shanghai
    Institute of Technical Physics, Chinese Academy of Sciences,
    200083 Shanghai, China}

\author{Zheng Tang}
\affiliation{Key Laboratory of Polar materials and Devices, Ministry of
  Education, East China Normal University, 200062 Shanghai, China}

\author{Junyu Zhu}
\affiliation{Key Laboratory of Polar materials and Devices, Ministry of
  Education, East China Normal University, 200062 Shanghai, China}

\author{Xiaodong Tang}
\affiliation{Key Laboratory of Polar materials and Devices, Ministry of
  Education, East China Normal University, 200062 Shanghai, China}

\author{Junhao Chu}
\email[Electronic mail: ]{jhchu@mail.sitp.ac.cn}
\affiliation{National Laboratory for Infrared Physics, Shanghai
    Institute of Technical Physics, Chinese Academy of Sciences, 200083
    Shanghai, China}
\affiliation{Key Laboratory of Polar materials and Devices, Ministry of
  Education, East China Normal University, 200062 Shanghai, China}

\begin{abstract}
In Hg$_{1-x}$Mn$_x$Te (x$\geq$0.16) monocrystal, the giant-diamagnetic (GDM) and
magnetization-step phenomena have been observed in spin glass (SG) regime. The
susceptibility of GDM is about 100-1000 times than that of classic diamagnetic.
It can be interpreted that: due to the long-range antiferromagnetic (AF) exchange
interactions and the non-uniform random distribution of Mn$^{2+}$ ions in
Hg$_{1-x}$Mn$_x$Te, a quasi-static spin wave forms and produces the GDM phenomenon
below the critical temperature and magnetic field. Meanwhile, this theory is proved
by Monte Carlo simulations in a two-dimensional AF cluster based on XY model.
Hence, it is possible to emerge long-range magnetic order structure in SG state.
\end{abstract}

\pacs{75.10.Nr, 75.20.-g, 75.30.Ds, 78.66.Hf, 75.50.Pp}

\maketitle

At low temperature, it is well known that the combined effects of randomness and
frustration may lead to spin-glass (SG) behavior in disordered spins system, such
as magnetic alloy, magnetic oxides and semimagnetic semiconductor (SMSC or DMS).
\cite{Binder:rmp-58-801-86,Gardner:rmp-82-53-10} For the SG of metallic alloy or
magnetic metal oxides, it is difficult to separate the contribution of the conduction
electrons from that of the localized spins , liking RKKY mechanism. Therefore,
for a better understanding of SG, Mn-based SMSCs are appropriate candidates for
studying SG in experiment, due to pure antiferromagnetic (AF) exchange interaction
and very low carriers concentration. In SG state, the global ground state of
system always is to a major concern problem and not be resolved until now.
Generally, mostly classical SG theories based on mean-field theory and short-range
AF exchange (e.g., Ising model and Sherrington-Kirkpatrick model\cite{Sherrington%
:prl-35-1792-75}) suppose that the spins have no long-range magnetic order but
instead have frozen or quasi-static orientations which vary randomly over macroscopic
distances at low temperatures.\cite{Binder:rmp-58-801-86} Meanwhile, it predicts
the limit concentration of the SG transition is about 17\% in SMSC with fcc
structure.\cite{Galazka:prb-22-3344-80}

Whereas, in Hg$_{1-x}$Mn$_x$Te (fcc structure), the existing results of magnetic
and specific heat experiments have proved that: (i) the AF exchange interaction
between Mn$^{2+}$ ions contains long-range exchange mechanisms, such as
Bloembergen-Rowland exchange;\cite{Jacek:DMS-1988,Brandt:AdvPhys-33-193-84,%
Lee:prb-37-8849-88} (ii) Mn$^{2+}$ ions are not exactly random uniform distribution
but random fluctuation distribution in space.\cite{Nagata:prb-22-3331-80,%
Mycielski:ssc-50-257-84,Anderson:prb-33-4706-86} Both features go against the basic
hypothesis of classical SG theories. Can it produce some new effects on the spin
arrangement of SG state, liking long-range magnetic ordered structures? From a
fundamental perspective, this is a very important issue in SG theory. For this
purpose, we investigate the magnetic properties of Hg$_{1-x}$Mn$_x$Te with variant
Mn concentrations, particularly near the SG regime.

In this work, the DC field susceptibility (2$-$300 K) and magnetization (2$-$10 K)
measurements with physical property measurement system (PPMS) of Quantum Design
were carried out on four Hg$_{1-x}$Mn$_x$Te monocrystal samples grown by modified
Bridgman method and annealed in Hg vapor. Four samples denoted as NO1 (x$\simeq$0.1),
NO2 (x$\simeq$0.16), NO3 (x$\simeq$0.191) and NO4 (x$\simeq$0.207) respectively,
where NO4 was the best monocrystal. Zero field cooled (ZFC) method was adopted
in the susceptibility measurements and the time interval for each measuring point
was 1 sec. In addition, Hall measurements proved four samples were strong p-type
and hole concentrations were lower than 10$^{14}$cm$^{-3}$ below 10 K.

\begin{figure}
\includegraphics*[angle=0,scale=1.0]{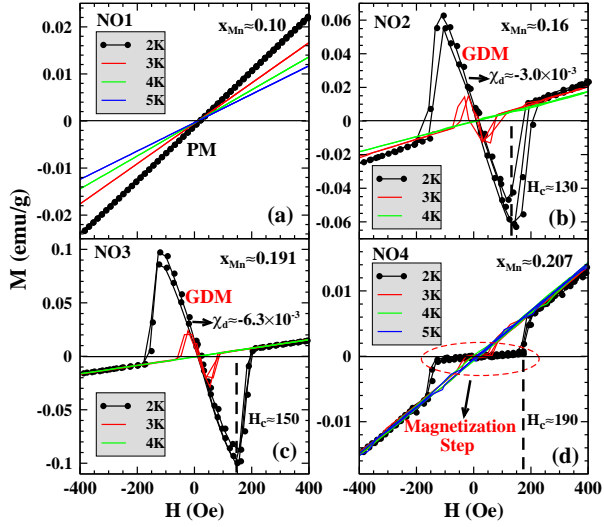}
\caption{\label{GDM-MH} The temperature-dependent magnetization curves of
  Hg$_{1-x}$Mn$_x$Te monocrystals with different Mn concentrations. These results
  show that when Mn concentration approaches or exceeds the limit of SG transition
  (17\%) in fcc structure, GDM and magnetization-step emerge below the critical
  temperature (T$_c$) and magnetic field (H$_c$). The susceptibility of GDM is
  about 100-1000 times than that of classic diamagnetic. In addition, the values
  of T$_c$ and H$_c$ go up as Mn concentration increases.}
\end{figure}
Figure\,\ref{GDM-MH} presents the magnetization curves of four samples at different
temperatures. From 2 K to 5 K, the magnetization curves of NO1 are simple straight
lines indicating good paramagnetic (PM) state, as shown in Figure\,\ref{GDM-MH}
(a). However, the remaining samples (NO2, NO3 and NO4) emerge novel and interesting
magnetization phenomena at low temperatures, as illustrated in Figure\,\ref{GDM-MH}(b),
(c) and (d). For NO2 and NO3, both of magnetization curves show giant-diamagnetic
(GDM) phenomenon below 4 K (called the critical temperature T$_c$). The following
are the master features of GDM: (1) The absolute value of GDM susceptibility is
very large and depends on Mn concentration ($\chi_d$=-3.0$\times$10$^{-3}$ for
NO2 and $\chi_d$=-6.3$\times$10$^{-3}$ for NO3), which is about 100$-$1000 times
greater than that of classic diamagnetic ($\chi_d$=-10$^{-6}\sim$-10$^{-5}$).
(2) There is a critical magnetic field (H$_c$) for the existence of GDM at firmed
temperature, e.g., H$_c\simeq$130 Oe for NO2 and H$_c\simeq$150 Oe for NO3 at 2 K.
As magnetic field exceeds H$_c$, GDM state rapidly changes into paramagnetic state.
(3) H$_c$ decreases with temperature rising. As regards NO4, when temperature
lower than 5 K (about the T$_c$), its magnetization curves emerge magnetization-step
instead of GDM. Meanwhile, this step also only exists under a critical magnetic
field (e.g., H$_c\simeq$190 Oe at 2 K), and gradually disappears as temperature
rises.

Comparing the results of magnetization measurements in four Hg$_{1-x}$Mn$_x$Te
samples, it is clear that: (i) When Mn concentration approaches or exceeds the
limit of spin-glass transition (17\%) in fcc structure, GDM and magnetization
-step will appear below the critical temperature (T$_c$) and magnetic field (H$_c$);
(ii) As Mn concentration increases, the T$_c$ and H$_c$ of GDM and magnetization
-step also slowly go up.

\begin{figure}
\includegraphics*[angle=0,scale=1.0]{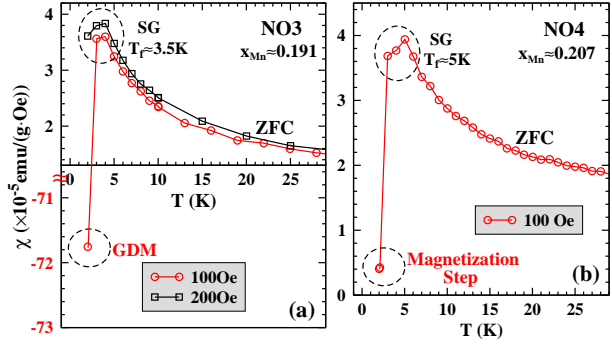}
\caption{\label{GDM-MT} The susceptibility curves ($\chi$-T) of NO3 and NO4
  measured by ZFC method. (a) shows the $\chi$-T curves of NO3 at 100 Oe and 200
  Oe. Both curves emerge a cusp structure at 3.5 K, which means SG transition.
  Meanwhile, at 100 Oe, the curve appears large negative value (GDM) below 3 K.
  (b) is the $\chi$-T curve of NO4 at 100 Oe which emerges both SG transition
  (5 K) and magnetization-step (below 3 K).}
\end{figure}
In order to making a further justification for GDM and magnetization-step, the
susceptibility measurements with the ZFC method were carried out on NO3 and NO4.
Figure\,\ref{GDM-MT} shows the susceptibility curves ($\chi$-T) of NO3 and NO4
from 2 K to 30 K at weak magnetic fields (H=100 Oe and 200Oe). For NO3, its
$\chi$-T curves emerge a cusp structure at about 3.5 K (exactly T$_c$), which
means the occurrence of SG transition, as shown in Figure\,\ref{GDM-MT}(a). In
the SG regime, the value of susceptibility markedly changes with magnetic field,
which are positive value at H=200 Oe (greater than H$_c$ at 2 K), but appears
large negative value (corresponding to GDM) at H=100 Oe (less than H$_c$ at 2 K).
As well, the $\chi$-T curve of NO4 at 200 Oe magnetic field also emerges SG
transition at 5 K (T$_c$) and the rapidly reduction of susceptibility below 5 K
(corresponding to magnetization-step), as illustrated in Figure\,\ref{GDM-MT}(b).
Hence, the results of susceptibility curves prove the existence of GDM and
magnetization-step again. More importantly, the GDM and magnetization-step are
associated with the SG transition.

Generally, the classical diamagnetic comes from electron orbit precession and
electromagnetic induction, the value of which is independent of temperature and
magnetic field. But the GDM of Hg$_{1-x}$Mn$_x$Te is related to temperature and
magnetic field. In addition, the relaxation magnetization model of SG proposed
by Lundgren, which assumes a uniform random distribution and short-range AF
interactions of magnetic ions, only leads to positive susceptibility (paramagnetic)
in DC magnetic measurements.\cite{Lundgren:jmmm-25-33-81,Lundgren:jpf-12-2663-82}
So, what are the reasonable physical mechanisms of GDM and magnetization-step
in Hg$_{1-x}$Mn$_x$Te?

We think these novel magnetization phenomena should come from the effects of
strong long-range AF interactions between Mn$^{2+}$ ions in Hg$_{1-x}$Mn$_x$Te.
The following are our reasons in details. In the magnetization process, whether
a spin of Mn$^{2+}$ can be flipped by magnetic field depends on the competition
of three factors: thermal fluctuation, magnetic field and AF interactions between
Mn$^{2+}$ ions. Usually, thermal fluctuation leads to spins chaotic flipping,
magnetic field causes the orderly arrangement of spins and AF interactions make
spins frozen. The condition of whether a spin is free or magnetic frozen is\cite{%
Twardowski:prb-36-7013-87}
\begin{equation}\label{AF-condition}
\sum_jJ_{\text{nn}}S_{\text{Mn}}^2\hat{s}_i\cdot\hat{s}_j\geq \frac{3}{2}k_BT
+g_{\text{Mn}}\mu_BS_{\text{Mn}}H
\end{equation}
where $\sum_jJ_{nn}S_{Mn}^2\hat{s}_i\cdot\hat{s}_j$ is the AF exchange energy
with the nearest neighbors, $\frac{3}{2}k_BT$ is the thermal kinetic energy and
g$_{Mn}$$\mu_B$S$_{Mn}$H is the magnetization energy.

\begin{figure}
\includegraphics[angle=0,scale=1.0]{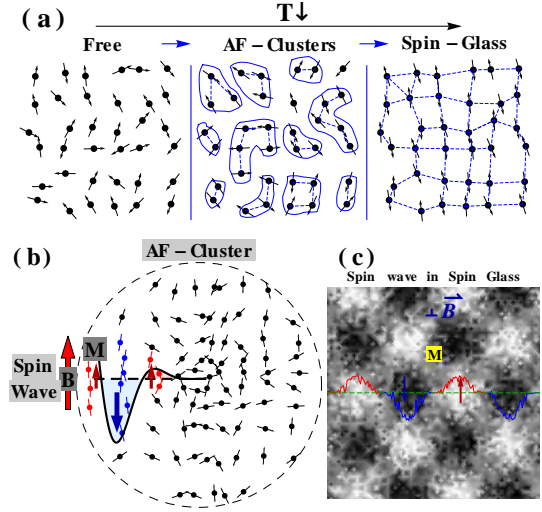}
\caption{\label{spin-cluster} (a) shows the formation of AF clusters and spin-glass
  as temperature decreases. (b) shows the structure of quasi-static spin wave in
  a 2D AF cluster with long-range AF interactions and non-uniform random distribution
  of Mn$^{2+}$ ions. (c) shows the possible interconnect structure of quasi-static
  spin waves in SG state.}
\end{figure}

At high temperatures, due to strong thermal fluctuations, Eq.\ref{AF-condition}
can not be satisfied even at zero magnetic field. Thus, the spins of Mn$^{2+}$
ions freely rotate and can be overturned easily by magnetic field. However, at
low temperatures, thermal fluctuations become weak and Eq.\ref{AF-condition} is
easy to satisfied, especially for high Mn concentration and strong long-range AF
interactions between Mn$^{2+}$ ions. In this case, the spins of Mn$^{2+}$ ions
are mostly frozen and form AF clusters, as shown in Figure\,\ref{spin-cluster}(a).

Then, according to Eq.\ref{AF-condition}, when Mn concentration with uniform random
distribution approaches or exceeds a critical value, it needs the same nonzero
magnetic field for all Mn$^{2+}$ ions to break down Eq.\ref{AF-condition} and
generate magnetization below a critical temperature (T$_c$). In other words,
magnetization can not appear when magnetic field and temperature are both less
than the critical values (H$_c$ and T$_c$) in high Mn concentration area, as shown
in NO4 sample. This is the physical mechanism of magnetization-step in Hg$_{1-x}$Mn$_x$Te.

As to the physical mechanism of GDM, it involves two key factors. One stems from
the effect of non-uniform random distribution of Mn$^{2+}$ ions in Hg$_{1-x}$Mn$_x$Te.
When the distribution of Mn$^{2+}$ is inhomogeneous in space, the magnetized
conditions (Eq.\ref{AF-condition}) of Mn$^{2+}$ ions will be different from each
other. It leads to, at the same temperature and magnetic field, the spins in high
concentration region are easy to frozen, but in low concentration region are easily
magnetized. As a result, the spin arrangement of Mn$^{2+}$ ions is also inhomogeneous
in space, especially in AF cluster which can be taken as the unit to compute the
spin arrangement of ground state in non-uniform random distributing spin system.

The other is due to the strong long-range AF interactions between Mn$^{2+}$ ions
in Hg$_{1-x}$Mn$_x$Te. In high Mn concentration area, this will cause that the
AF exchange energy with further neighbors (J$_{fn}$(R)S$_{Mn}^2$) is stronger
than thermal kinetic energy of Mn$^{2+}$ ion when temperature below a critical
value. Consequently, the spins of Mn$^{2+}$ ions emerge multi-frustration effect,
which exist not only between the nearest neighbors but also with further neighbors.
strongly correlated

Combining the effects of two factors, a spin flip of Mn$^{2+}$ ion occurred at
the edge of AF cluster can impact the spin arrangement of both the nearest
neighbors and further neighbors, and then produce a series of chain reactions in
AF cluster, liking "dominoes" effect. At the right temperature and magnetic field,
this effect can create a quasi-static spin wave with oscillating spin arrangement,
as presented in Figure\,\ref{spin-cluster}(b). More importantly, the amplitude
of spin waves also depends on the Mn concentration, which may lead to a net
negative magnetic moment (opposite to magnetic field direction) in non-uniform
random distributing AF clusters. Meanwhile, in the SG state, AF clusters are
interconnect with each other forming a AF "super-cluster". Hence, the quasi-static
spin waves can propagate in space, causing the GDM phenomenon in Hg$_{1-x}$Mn$_x$Te,
as shown in Figure\,\ref{spin-cluster}(c). This quasi-static spin wave caused by
local spins is similar to the spin-density wave (SDW) caused by electrons in
chromium alloys.\cite{Fawcett:rmp-66-25-94}

It is hard to strictly confirmed the model of quasi-static spin waves by analytical
method. However, it is possible to verify the rationality of this model by numerical
simulation. For this purpose, Monte Carlo (MC) simulation was employed to compute
the spin configuration of a two dimensional (2D) AF cluster based on XY model at
different temperatures and magnetic fields. Moreover, in order to overcoming the
influence of metastable as much as possible, simulated annealing was applied to
seek the ground state of 2D AF cluster.\cite{William:Numerical-2007}

As an example, the simulation results of a 2D AF cluster referring to the properties
of Hg$_{809}$Mn$_{0.191}$Te (NO3 sample) are discussed in this paper.\cite{%
Rogalski:infp-31-117-91,HgMnTe:LBmanul} This 2D AF cluster is taken as circular
shape (the diameter is about 32.52 nm) and contains 399 Mn$^{2+}$ ions with
non-uniform random distribution along the radial direction. In the central region,
the average distance between the nearest Mn$^{2+}$ ions ($\bar{a}_{nn}$) is about
1.163 nm. From the center to the edge, $\bar{a}_{nn}$ decreases linearly, which
is about 1.221 nm in the edge region. In Hg$_{1-x}$Mn$_x$Te, the exchange function
of long-range AF interactions between Mn$^{2+}$ ions is taken as
$J\left(R_{ij}\right)=J_1\left(\bar{a}_{nn}/R_{ij}\right){}^5$, where R$_{ij}$ is
the distance between two Mn$^{2+}$ ions and J$_1$(=-7.0K$_B$) is AF exchange
integral constant between the nearest neighbors Mn$^{2+}$ ions.\cite{Twardowski:%
prb-36-7013-87,Spalek:prb-33-3407-86} The Hamiltonian of each Mn$^{2+}$ ion is
in form of Eq.\,\ref {Hamiltonian}. In MC simulations, the truncation radius
R$_{cut}$=5$\bar{a}_{nn}$, S$_{Mn}$=5/2, g$_{Mn}$=2, and the number of MC steps
is 10$^5$ at each temperature point.
\begin{equation}\label{Hamiltonian}
H_i=-\sum _{R_{\text{ij}}\leq R_{\text{cut}}} J\left(R_{\text{ij}}\right)
S_{\text{Mn}}^2\hat{s}_i\cdot \hat{s}_j-g_{\text{Mn}}\mu _B\overset{\to }{B}
\cdot \overset{\to }{S}_i
\end{equation}

\begin{figure}
\includegraphics[angle=0,scale=1.0]{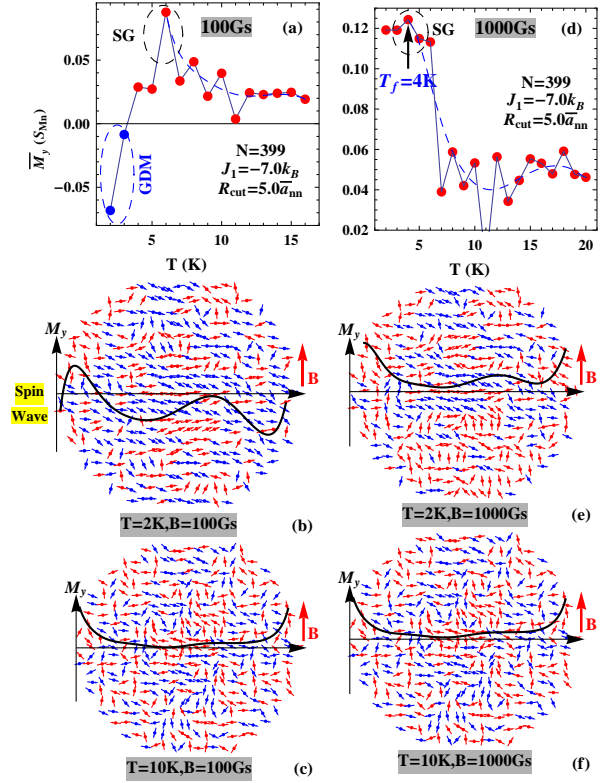}
\caption{\label{MC-results} The typical MC simulation results of a 2D AF cluster
  with long-range AF interactions and non-uniform random distribution. At 100 Gs,
  the $\bar{M}_y$-T curve emerges SG transition (about 5 K) and GDM (below 4 K)
  for (a). Meanwhile, the spin configurations of ground state show a quasi-static
  spin wave which changes with temperature for (b) and (c). However, at 1000 Gs,
  the $\bar{M}_y$-T curve only emerges SG transition and the quasi-static spin
  wave is faint in the 2D AF cluster, for (d), (e) and (f).}
\end{figure}
Figure\,\ref{MC-results} show the typical MC simulation results of the 2D AF
cluster, including the $\bar{M}_y$-T curves ($\bar{M}_y$ denotes the average
magnetic moment of all Mn$^{2+}$ ions along magnetic field direction) and the
spin configurations of ground state at different temperatures and magnetic fields.
When magnetic field is 100 Gs, the $\bar{M}_y$-T curve emerges both cusp structure
(SG transition) and negative value (paramagnetic-diamagnetic transition below 4 K),
moreover, the spin configurations of ground state prove that a quasi-static spin
wave really exists and gradually disappears as temperature rises, as shown in
Figure\,\ref{MC-results}(a), (b) and (c). However, when magnetic field increases
to 1000 Gs, the $\bar{M}_y$-T curve only emerges cusp structure, and the
quasi-static spin wave is faint or absent which is insufficient to cause
paramagnetic-diamagnetic transition at low temperatures, as shown in
Figure\,\ref{MC-results}(d), (e) and (f).

These MC simulation results of the 2D AF cluster demonstrate that: a quasi-static
spin wave really exists and leads to paramagnetic-diamagnetic transition (similar
to the GDM in experiments) below the critical temperature, then gradually vanishes
as temperature or magnetic field increases. Meanwhile, these results are basically
consistent with the experimental results of GDM in Hg$_{1-x}$Mn$_x$Te (NO2 and NO3
samples). Hence, in Hg$_{1-x}$Mn$_x$Te, the model of quasi-static spin waves
inducing GDM is reasonable.

In conclusion, the results of DC magnetic measurements prove Hg$_{1-x}$Mn$_x$Te
($x\geq 0.16$) monocrystals emerge giant-diamagnetic (GDM) and magnetization-step
when temperature and magnetic field below the critical values (T$_c$ and H$_c$).
The susceptibility of GDM is about 100-1000 times than that of classic diamagnetic.
These novel magnetic phenomena come from the effects of long-range AF exchange
interactions and non-uniform random distribution of Mn$^{2+}$ ions in Hg$_{1-x}$Mn$_x$Te,
such as inducing a quasi-static spin wave which produces the GDM. In addition,
in a 2D AF cluster with long-range AF interactions and non-uniform random
distribution, Monte Carlo simulations confirm that quasi-static spin waves really
exist, which lead to paramagnetic-diamagnetic transition in the SG regime and
gradually disappear as temperature or magnetic field increases. Hence, it is
possible to emerge long-range magnetic order structure in SG state.

The authors thank Wanqi Jie and Tao Wang of Northwestern Polytechnical University
for the preparation of the HgMnTe monocrystal sample. This work was sponsored by
the STCSM (09JC1415600), the NSF (10927404, 60723001 and 60821092) and the NBRP
(2007CB924902) of China.



\end{document}